\documentstyle[epsfig]{mn}

\newif\ifAMStwofonts

\ifoldfss
  \ifCUPmtlplainloaded \else
    \NewTextAlphabet{textbfit} {cmbxti10} {}
    \NewTextAlphabet{textbfss} {cmssbx10} {}
    \NewMathAlphabet{mathbfit} {cmbxti10} {} 
    \NewMathAlphabet{mathbfss} {cmssbx10} {} 
  \fi
  \ifAMStwofonts
    \ifCUPmtlplainloaded \else
      \NewSymbolFont{upmath} {eurm10}
      \NewSymbolFont{AMSa} {msam10}
      \NewMathSymbol{\upi}     {0}{upmath}{19}
      \NewMathSymbol{\umu}     {0}{upmath}{16}
      \NewMathSymbol{\upartial}{0}{upmath}{40}
      \NewMathSymbol{\leqslant}{3}{AMSa}{36}
      \NewMathSymbol{\geqslant}{3}{AMSa}{3E}

    \fi
  \fi
\fi 

\ifnfssone
  \newmathalphabet{\mathit}
  \addtoversion{normal}{\mathit}{cmr}{m}{it}
  \addtoversion{bold}{\mathit}{cmr}{bx}{it}
  \newmathalphabet{\mathbfit} 
  \addtoversion{normal}{\mathbfit}{cmr}{bx}{it}
  \addtoversion{bold}{\mathbfit}{cmr}{bx}{it}
  \newmathalphabet{\mathbfss} 
  \addtoversion{normal}{\mathbfss}{cmss}{bx}{n}
  \addtoversion{bold}{\mathbfss}{cmss}{bx}{n}
  \ifAMStwofonts
    \ifCUPmtlplainloaded \else
      \UseAMStwoboldmath
      \makeatletter
      \new@mathgroup\upmath@group
      \define@mathgroup\mv@normal\upmath@group{eur}{m}{n}
      \define@mathgroup\mv@bold\upmath@group{eur}{b}{n}
      \edef\UPM{\hexnumber\upmath@group}
      \new@mathgroup\amsa@group
      \define@mathgroup\mv@normal\amsa@group{msa}{m}{n}
      \define@mathgroup\mv@bold\amsa@group{msa}{m}{n}
      \edef\AMSa{\hexnumber\amsa@group}
      \makeatother
      \mathchardef\upi="0\UPM19
      \mathchardef\umu="0\UPM16
      \mathchardef\upartial="0\UPM40
      \mathchardef\leqslant="3\AMSa36
      \mathchardef\geqslant="3\AMSa3E
    \fi
  \fi
\fi 

\ifnfsstwo
  \DeclareMathAlphabet{\mathbfit}{OT1}{cmr}{bx}{it}
  \SetMathAlphabet\mathbfit{bold}{OT1}{cmr}{bx}{it}
  \DeclareMathAlphabet{\mathbfss}{OT1}{cmss}{bx}{n}
  \SetMathAlphabet\mathbfss{bold}{OT1}{cmss}{bx}{n}
  \ifAMStwofonts
    \ifCUPmtlplainloaded \else
      \DeclareSymbolFont{UPM}{U}{eur}{m}{n}
      \SetSymbolFont{UPM}{bold}{U}{eur}{b}{n}
      \DeclareSymbolFont{AMSa}{U}{msa}{m}{n}
      \DeclareMathSymbol{\upi}{0}{UPM}{"19}
      \DeclareMathSymbol{\umu}{0}{UPM}{"16}
      \DeclareMathSymbol{\upartial}{0}{UPM}{"40}
      \DeclareMathSymbol{\leqslant}{3}{AMSa}{"36}
      \DeclareMathSymbol{\geqslant}{3}{AMSa}{"3E}
    \fi
  \fi
\fi 

\ifCUPmtlplainloaded \else
  \ifAMStwofonts \else 
    \def\upi{\pi}
    \def\umu{\mu}
    \def\upartial{\partial}
  \fi
\fi

\title[H1320+551: Sy 1.8/1.9 unabsorbed AGN]{H1320+551: A Seyfert
1.8/1.9 galaxy with an unabsorbed  X-ray spectrum\protect{\thanks{Based partly
on observations obtained with XMM-Newton, an ESA science mission with
instruments and contributions directly funded by ESA member states and
the USA (NASA)}}}
\author[X. Barcons, F.J. Carrera, M.T. Ceballos]
       {X. Barcons, F.J. Carrera, M.T. Ceballos \\
        Instituto de F\'\i sica de Cantabria (CSIC-UC), 39005 Santander, Spain}
\date{September 2002}

\pagerange{\pageref{firstpage}--\pageref{lastpage}}
\pubyear{2002}

\begin{document}

\maketitle

\label{firstpage}

\begin{abstract}
We present new optical spectroscopic and XMM-Newton X-ray observations
of the Active Galactic Nucleus H1320+551. The optical data (consistent
with but of better quality than a previously published spectrum) show
this source to be a Seyfert 1.8/1.9 galaxy at $z=0.0653$. The narrow
line region is significantly reddened, with a Balmer decrement
$H\alpha/H\beta\sim 6$ and the broad line region, with a barely
detectable H$\beta$ broad component, shows a much pronnounced Balmer
decrement ($H\alpha/H\beta> 27$). In spite of this, the EPIC-pn
X-ray spectrum exhibits a power-law continuum with a soft excess that
is well fitted by a black body, with no photoelectric absorption above
the galactic value. A Fe K emission line is also seen at a rest-frame
energy $\sim 6.5$~keV with an equivalent width of $\sim 400$ eV, far
too weak for the source being Compton-thick. Reconciling the optical
and X-ray data requires the narrow line region being internally
reddened but with small covering factor over the nuclear emission and
the Balmer decrement of the broad line region being an intrinsic
property rather than caused by reddening/absorption. The H1320+551
Seyfert 1.8/1.9 galaxy is not consistent with being an obscured Seyfert 1
nucleus, i.e., it does not match the basic AGN unified scheme
hypothesis. 

\end{abstract}

\begin{keywords}
galaxies:active, Seyfert; X-rays:galaxies
\end{keywords}

\section{Introduction}

The X-ray spectral properties of Active Galactic Nuclei (AGN) are
generally well correlated with their optical appearance.  Seyfert 1
galaxies and QSOs have usually steep X-ray spectra with little, if
any, intrinsic photoelectric absorption (Nandra \& Pounds 1994).
Seyfert 2 galaxies have, on the contrary, absorbed X-ray spectra
(Smith \& Done 1996).  This is hardly surprising in the framework of
the simplest version of the AGN unified model (Antonucci 1993), as the
molecular gas and dust that prevents the direct view of the Broad-Line
region in type 2 Seyferts (the ``torus'') is likely to also contain
atomic gas that will absorb soft X-rays.

Unified AGN models for the cosmic X-ray background (XRB), first
suggested by Setti \& Woltjer (1989) and later worked out by Madau,
Ghisellini \& Fabian (1994), Comastri et al (1995) and Gilli, Salvati
\& Hasinger (2001), make use of this feature.  A broad distribution of
photoelectric absorbing columns is assumed, which results in an
integrated XRB with the required spectral shape. X-ray surveys are
expected to reveal mostly type 1 AGN for soft unabsorbed X-ray sources
and type 2 AGN for hard absorbed X-ray sources.  Indeed, soft X-ray
surveys carried out with $ROSAT$ are rich in type 1 AGNs and QSOs
(see, e.g., Mason et al 2000 and Lehmann et al 2001 for medium and
deep $ROSAT$ surveys). On the contrary, hard X-ray surveys carried out
with $BeppoSAX$ do contain large numbers of type 2 AGN (Fiore et al
1999).

Risaliti, Maiolino \& Salvati (1999) found that the X-ray absorption
in a sample of [OIII]-selected type 1.8, 1.9 and 2 AGN reveals much
higher absorbing columns than in samples selected by other means, a
large fraction of the sources being actually Compton-thick. Since
[OIII] emission is supposed to arise above the obscuring torus, [OIII]
selection is likely to be an orientation-independent measure of the
AGN intrinsic luminosity. Still, a rough trend of increasing X-ray
absorption with AGN Seyfert type is found (Risaliti, Maiolino \&
Salvati 1999, Alonso-Herrero et al 1997).

However, a number of studies show that the optical obscuration to
X-ray absorption relation is not as simple as predicted by the AGN
unified model.  A hard spectrum selection of $ROSAT$ X-ray sources,
which was supposed to favour absorbed X-ray sources
(Page et al 2000), revealed mostly type 1 AGN, while more type 2 AGN
were expected (Page et al 2001). Granato, Danese \& Franceschini
(1997) proposed that X-ray absorption takes place mostly in dust-free
regions, below the dust sublimation radius. This will accomodate the
existence of X-ray absorbed, optically unobscured type 1 AGN.

Pappa et al (2001) have found, in a sample of Seyfert 2 galaxies, some
extreme cases without or with very small apparent intrinsic X-ray
absorption. The apparent lack of X-ray absorption has been attributed
to some of these Seyfert 2 galaxies being actually Compton-thick, in
which case we would be witnessing only scattered radiation and host
galaxy emission from a circumnuclear starburst below 10 keV.  Bassani
et al (1999) propose a three-dimensional diagnostic that would
discriminate between that possibility and true lack of absorption, on
the basis of a diagram displaying the equivalent width of the Fe K
emission line versus the transmission $T$ defined as the ratio between
the 2-10 keV X-ray flux (supposed to measure the emission transmitted
through the torus) and the reddening-corrected [OIII] flux, assumed to
measure the intrinsic AGN emission.  Compton-thick type 2 AGN lie
invariably at the high equivalent width low transmission end. That has
helped to unmask a number of puzzling Seyfert 2 galaxies apparently
unabsorbed.  The data quality of the $ASCA$ spectrum used by Pappa et
al (2001) was certainly good enough to discard a Compton-thick origin,
but did not enable the discrimination between a dusty warm absorber
and a genuine broad-line region free AGN.

In this paper we present new optical and X-ray observations of
H1320+551. This source was discovered by HEAO-1 as part of the
Modulation Collimator-Large Area Sky Survey (Wood et al 1984).  The
inferred 2-10 keV flux was $\sim 2\times 10^{-11}\, {\rm erg}\, {\rm
cm}^{-2}\, {\rm s}^{-1}$ (Ceballos \& Barcons 1996), although
confusion could be an issue. Remillard et al (1993) identified this
X-ray source with a type 1 AGN at $z=0.064$ ($RA=13^h22^m49.2^s$,
$DEC=+54^{\circ}55'28''$) . However, their optical spectrum was noisy
and of poor spectral resolution ($\sim 10$\AA) with a barely visible
H$\beta$ line.

H1320+551 was also detected in the $ROSAT$ all-sky survey and
identified in the $ROSAT$ Bright Survey (Schwope et al 2000). The
$ROSAT$ PSPC count rate was $0.23\pm 0.024$ ct/s, corresponding to an
absorption-corrected 0.5-2 keV flux of $\sim (1.9\pm 0.2)\times
10^{-12}\, {\rm erg}\, {\rm cm}^{-2}\, {\rm s}^{-1}$. The PSPC
Hardness Ratio was $+0.12\pm 0.10$ . The source had therefore a
moderately steep soft X-ray spectrum in the $ROSAT$ band, but the
small flux ratio $S(0.5-2)/S(2-10)$ (if the 2-10 keV flux was correct)
suggested absorption. $ASCA$ observed H1320+551 in May 1999 and we
analyze the archival data here.

In what follows we present new optical spectroscopy and XMM-Newton
X-ray observations of H1320+551. Our optical data (obtained in 1998)
shows this source to be a Seyfert 1.8/1.9 galaxy at $z=0.0653$ with a
significantly reddened narrow line region (section 2). The XMM-Newton
X-ray spectrum is well described by a type-1 like continuum (power-law
plus black body) with no intrinsic photoelectric absorption, plus a Fe
K emission line complex (section 3).  In section 4 we show that the
unabsorbed X-ray spectrum is inconsistent with a
dust-reddening origin for the large Balmer decrement in the broad line
region, which is instead more likely to be intrinsic to the broad-line
clouds, contrary to the predictions of the AGN unified model.

\section{Optical observations}

H1320+551 was observed by the 4.2m William Herschel Telescope at the
Observatorio del Roque de Los Muchachos in the island of La Palma
(Canary Islands, Spain), on February 26,  1998.  We used the
ISIS double spectrograph with 600 line/mm gratings on both the blue
and red arms, with the wavelengths centered at 5200 and 7000 \AA\
respectively, in order to observe the H$\beta$+[OIII] region in the
blue and the H$\alpha$+[NII]+[SII] in the red.  During the second half
of the night, when this observation was performed, the sky was clear
and probably photometric.  The seeing was $\sim 1.5$ arcsec and we
used a 1.5 arcsec slit width. Observations were carried out with the
slit aligned to paralactic angle. Two exposures of 300 sec each were
taken.  One of the blue spectra had a cosmic ray hit on top of the 
H$\beta$ line and we have not used it.  The two red-arm exposures were 
co-added.

Reduction and calibration was performed according to a standard
sequence under the IRAF package. It included de-biasing,
flat fielding, wavelength calibration with arc lamps, and flux
calibration using a standard star and standard extinction curve for
the observatory. The wavelength calibration gave residuals of 0.05 and
0.02 \AA\ in the blue and in the red respectively.  The measured
spectral resolution (using gaussian fits to unblended arc lines) gave
2.22 \AA\ and 2.16 \AA\ at the central wavelengths of the blue and red
channels respectively. Given the relatively poor seeing, the
spectrophotometric calibration is far from accurate, but certainly
good to within a factor of 2. 

\begin{figure}
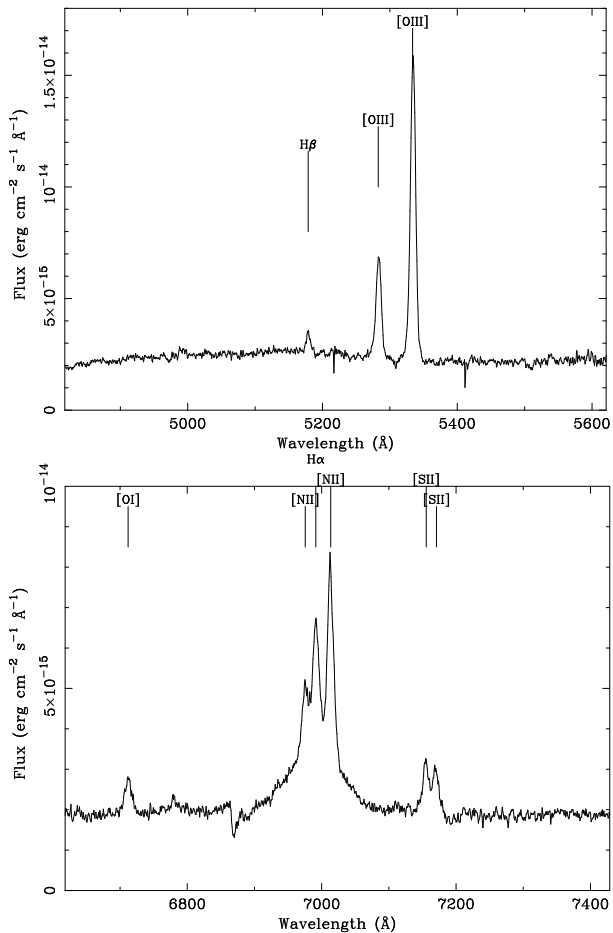

\epsfig{file=H1320+551_blue.ps,height=8cm,angle=270}
\epsfig{file=H1320+551_red.ps,height=8cm,angle=270}
\caption{Optical spectrum of H1320+551, blue (top) and red (bottom) arms.}
\label{fig-optical-spectra}
\end{figure}

Fig \ref{fig-optical-spectra} shows the resulting spectra in the blue
and the red arms with marks on the most important lines. It is clear
that the H$\beta$ line is weak and dominated by a narrow component,
with very weak, if any, broad component.  The H$\alpha$+[NII] blend
does, on the contrary, exhibit both a narrow and a broad H$\alpha$
component.   Therefore, according to the standard
classification by Osterbrock (1981), the source is in principle a
Seyfert 1.9.  Remillard et al (1993) classified it as a Seyfert 1 on
the basis of a low spectral resolution ($\sim 10$\AA) low signal to
noise optical spectrum (see fig 3 in that paper). In their spectrum
the H$\beta$ line was barely visible and the H$\alpha$+[NII] complex
could not be deblended.

\begin{figure}
\epsfig{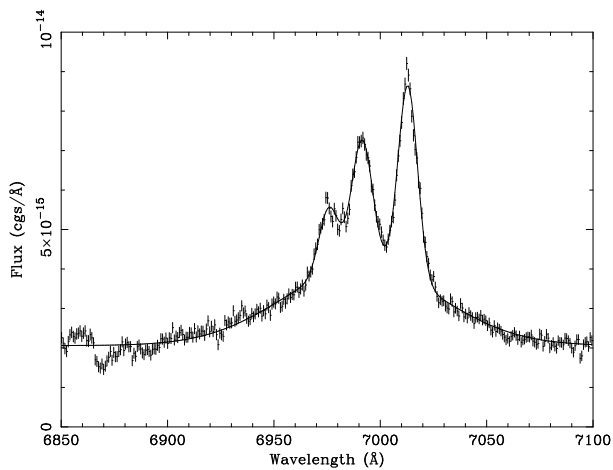}
\caption{Detail of the optical spectrum and fit to the
H$\alpha$+[NII]complex (see text for details)}
\label{fig-hanii}
\end{figure}

In order to measure line intensities, we selected a small portion of
the spectrum around each complex and we fitted the spectrum (via
$\chi^2$ minimisation, using the {\tt qdp} fitting routines) with a
constant plus a gaussian for each putative line.  Errors on the
intensity of each individual wavelength channel were propagated during
the reduction process, in order to perform this fit correctly. The
H$\beta$ and each one of the [OIII] doublet lines were fit
independently.  The [SII] doublet, slightly blended, was fitted
simultaneously with two gaussians at the same redshift and with the
same width. In the case of the $H\alpha$+[NII] blend, the 3 lines were
simultaneously fitted to gaussians, all of them with the same redshift
and with the additional constraint that the two [NII] lines had the
same width. An additional broad H$\alpha$ line had to be added to
achieve a good fit. The result is shown in fig~\ref{fig-hanii}. Based
on the existence of a broad H$\alpha$ component, we searched for a
corresponding H$\beta$ one with the same velocity width.  This
constrained fit yields a ``detection'' of a weak broad H$\beta$ component
that is not required if the line width is left as a free
parameter. This is the reason of the uncertainty in the 1.8/1.9
Seyfert type classification of this source.

Table \ref{Tab-optical-lines} lists the measured line intensities and
velocity widths (FWHM). These were computed from the gaussian fits,
subtracting in quadrature the spectral resolution of the corresponding
spectrograph channel ($125\, {\rm km}\, {\rm s}^{-1}$ and $100\, {\rm
km}\, {\rm s}^{-1}$ in the blue and red arms respectively). The
redshift that we fit to the emission lines is $z=0.06531\pm 0.00002$,
refining the value $z=0.064$ reported by Remillard et al (1993).

\begin{table}
\label{Tab-optical-lines}
\begin{tabular}{l c c}
Line   & Line flux & FWHM \\
       & ($\times 10^{-15}\, {\rm erg}\, {\rm cm}^{-2}\, {\rm s}^{-1}$) & (${\rm km}\, {\rm
s}^{-1}$)\\

\hline

H$\beta 4861$ (narrow) & 6.6  &   400 \\
H$\beta 4861$ (broad) & 6.7  &   3710$^*$ \\
$[OIII] 4958$   & 45.5 &   525 \\
$[OIII] 5007$   & 130.4 &  500  \\
$[OI] 6300$ & 13.5 & 520\\
$[NII] 6548$    & 19.1  &   450\\
H$\alpha$ 6562 (narrow) & 40.0 & 490 \\
H$\alpha$ 6562 (broad)  & 182.9 &  3710\\
$[NII] 6583$    & 56.4  &   450\\
$[SII] 6716$    & 15.7  &   445\\
$[SII] 6730$    & 14.4  &   445\\
\hline
\end{tabular}
\caption{Measured line fluxes and velocity widths, as fitted from
gaussians. A velocity of $125\, {\rm km}\, {\rm s}^{-1}$ and of $100\, 
{\rm km}\, {\rm s}^{-1}$ has been subtracted in quadrature in the blue 
and red arm lines respectively, to account for the spectrograph
resolution. The H$\beta$ broad line has been fit with the velocity
fixed ($^*$) at the value found for H$\alpha$, as otherwise this feature
is not detected.}
\end{table}

From the line intensities we see that this AGN has significant
reddening. The Balmer decrement of the Narrow Line Region (NLR), as
traced by the narrow line components, is $(H\alpha/H\beta)_{\rm
NLR}\sim 6$ (corresponding to $E(B-V)\sim 0.5^{mag}$), where $\sim 3$
is expected for a variety of models under case B recombination and
optically thin NLR gas. Narrow line ratios can be reddening corrected
using the Balmer decrement as the indicator. Following Baldwin,
Phillips \& Terlevich (1981) we find $\log [[OIII]5007/H\beta
4861]=1.20$ and $\log [[NII]6583/H\alpha 6562]=0.145$ which, as
expected, place this object in the AGN zone in line diagnostic
diagrams (e.g. Osterbrock 1989, fig 12.1). We have further used the
measured Balmer decrement from the narrow lines ($H\alpha/H\beta\sim
6$), to estimate a gas column density of $N_{H}({\rm NLR})\sim
3\times10^{21}\, {\rm cm}^{-2}$, assuming standard gas-to-dust ratio
(Bohlin et al 1978).

Since the [OIII] emission is likely to come from well above the torus,
the intensity of the [OIII]5007 line is supposed to be an
orientation-independent estimator of the total AGN power.  Following
Bassani et al (1999) and Pappa et al (2001), who use the interstellar
reddening law by Savage \& Mathis (1979), we estimate the unreddened
[OIII]5007 flux by correcting the observed one by a factor
$[(H\alpha/H\beta)_{\rm NLR}/3]^{2.94}$. The resulting [OIII]5007 flux
is $\sim 1.34\times 10^{-12}\, {\rm erg}\, {\rm cm}^{-2}\, {\rm
s}^{-1}$

When a similar analysis is performed in the Broad Line Region (BLR), a
Balmer decrement $(H\alpha/H\beta)_{\rm BLR}\sim 27$ is found. This
should be considered as a lower limit, as the existence of an H$\beta$
broad component (i.e. the 1.8 Seyfert character of H1320+551) can only
be established via constrained parameter fitting. If this Balmer
decrement is interpreted in terms of reddening over a standard value
of 3, a value of $E(B-V)_{\rm BLR}\sim 2^{mag}$ is found, which  for a
standard dust to gas ratio corresponds to a H column density of
$N_{H}({\rm BLR})\sim 10^{22}$.  In the framework of the AGN unified
model the difference between optical spectroscopic Seyfert types is
due to an orientation effect which results in both reddening of the
BLR and absorption of X-rays. Under these circumstances the above
absorption column density should be seen in the X-ray data.

We have also tried to fit the broad band optical spectra obtained here
with a model mixing a reddened QSO, from the Francis et al (1991)
template and a E/S0 galaxy template, from the Coleman, Wu \& Weedman
(1980)  model. A good
simultaneous description of the blue and red spectrum is achieved with
the QSO template, which can accomodate a very small amount of
reddening ($E(B-V)<0.1^{mag}$) but it certainly needs some host
galaxy light.  The latter contributes about 40-60 per cent of the optical
spectrum at our reference point at 5550 \AA .  Significantly larger
reddening over the Francis et al (1991) QSO template simply does not
match the data.  That suggests that whatever causes the large BLR
Balmer decrement does not appear to be nuclear reddening.

\section{Previous X-ray observations}

\subsection{HEAO-1}

As already mentioned, X-ray emission from H1320+551 was discovered in
the MC-LASS survey, which assigned it a count rate of $(4.1\pm
0.8)\times 10^{-3}$ ct/s. Observations were carried out during 1977.
Ceballos \& Barcons (1996) assumed a $\Gamma=1.7$ power law spectrum
and computed a 2-10 keV flux of $2.0\times 10^{-11}\, {\rm erg}\, {\rm
cm}^{-2}\, {\rm s}^{-1}$.

\subsection{$ROSAT$}

H1320+551 was detected in the $ROSAT$ all-sky survey (1990/91) as source 1RXS
J132248.5+545526 (Schwope et al 2000) with a PSPC count rate of
$0.23\pm 0.024$ ct/s, and a Hardness Ratio
\[
HR_{PSPC}={C(0.5-2.0)-C(0.1-0.4)\over C(0.5-2.0)+C(0.1-0.4)}=0.12\pm
0.10
\]
where $C(0.1-0.4)$ and $C(0.5-2.0)$ are the PSPC counts collected in
the 0.1-0.4 keV and 0.5-2.0 keV bands respectively. The 0.5-2.0 keV
absorption-corrected flux was computed in Ceballos \& Barcons (1996)
by assuming a $\Gamma=1.9$ power law spectrum with galactic
absorption, resulting in $(1.9\pm 0.2)\times 10^{-12}\, {\rm erg}\,
{\rm cm}^{-2}\, {\rm s}^{-1}$.

\subsection{ASCA}

\begin{figure}
\epsfig{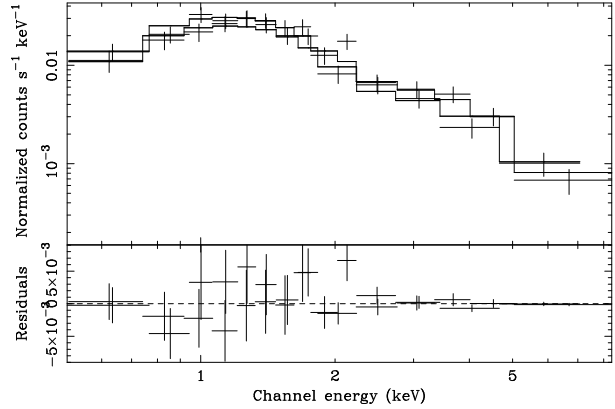}
\caption{ASCA SIS0 and SIS1 spectrum of H1320+551 together with
power-law fit plus Galactic absorption.}
\label{fig-ascaspec}
\end{figure}

$ASCA$ observed H1320+551 on May 10 of 1999 for 10 ks. We have retrieved
the pipeline-reduced data from the HEASARC public archive.  The SIS0
and SIS1 data over the 0.5-8 keV band can be well fitted with a single
power law with galactic absorption, resulting in a $\chi^2=17.4$ for
26 degrees of freedom (see fig. \ref{fig-ascaspec} for the $ASCA$
spectrum). The data do not require additional absorption or a soft
excess.

\begin{figure}
\epsfig{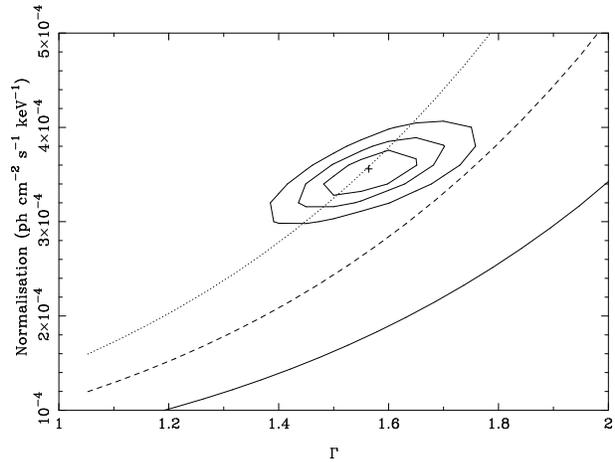}
\caption{Confidence contours (1,2 and 3 sigma for 2 parameters) for
the power law index and normalisation of the fit to the ASCA data for
H1320+551. Lines at various 2-10 flux levels are also overlayed: 1.0
(continuous), 1.5 (dashed) and 2.0 (dotted) $\times 10^{-12}\, {\rm
erg}\, {\rm cm}^{-2}\, {\rm s}^{-1}$}
\label{fig-cont-asca}
\end{figure}

Fig~\ref{fig-cont-asca} shows the confidence contours for the
power-law photon index ($\Gamma=1.53\pm0.1$) and the normalisation
required by the $ASCA$ data. We have also overlayed lines with various
2-10 keV flux levels, for comparison with the XMM-Newton observations.
Indeed the $ASCA$ 2-10 keV flux is 10 times smaller than the flux
assigned to this source by the MC-LASS survey, although part of this
discrepancy might be due to source confusion in the HEAO-1 data.

\section{XMM-Newton X-ray observations}

H1320+551 was observed by XMM-Newton (Jansen et al 2001) during
revolution 366 on the 8th of December 2001. This observation was taken
as part of the Guaranteed Time of the Survey Science Centre. The EPIC
pn camera (Str\"uder et al 2001) was operated in small window mode
with the Thin filter. The MOS cameras (Turner et al 2001) were
operated in timing mode. The RGS spectrographs (den Herder et al 2001)
were operated in standard spectroscopy mode, but our target produced a
very faint signal. Due to the brightness of the source in the optical,
the OM optical camera (Mason et al 2001) was switched off during this
observation. In this paper we present the analysis of the EPIC pn data
only, as the Science Analysis Software (SAS) version 5.3.3 provides full
support for calibration matrices of the small window mode.  The same
version of the SAS is not meant to provide response matrices for the
MOS cameras in timing mode.

The EPIC pn exposure time was 20 ks, with a count rate of 1.7 cts/s
in the 0.2-12 keV band. The background did not flare significantly
during the observation. We used the EPIC pn calibrated event list
provided in the pipeline products, which were obtained by processing
the observation data file with SAS version 5.2.  To gain full support
in the analysis, we extracted the source and background spectrum, and
generated re-distribution matrices and ancillary response files using
SAS version 5.3.3. The spectrum was also grouped in bins containing a
minimum of 30 counts. All counts outside the 0.2-12. keV range were
ignored.

As a first exercise, we computed the hardness ratios
\[
HR_1={S(2.0-4.5)-S(0.5-2.0)\over S(2.0-4.5)+S(0.5-2.0)}\approx -0.681
\]

\[
HR_2={S(4.5-12.0)-S(2.0-4.5)\over S(4.5-12.0)+S(2.0-4.5)}\approx
-0.343
\]
These turn out to be entirely consistent with the average hardness
ratios of the broad-line AGNs found in the AXIS medium sensitivity
survey: $ \langle HR_1\rangle =-0.68\pm 0.01$ and $\langle
HR_2\rangle=-0.31\pm 0.04$ (Barcons et al 2002). The type 2 AGN in
that survey have an average $\langle HR_1\rangle=-0.48\pm 0.11$, which appears
marginally harder than the values for H1320+551.

\begin{figure}
\epsfig{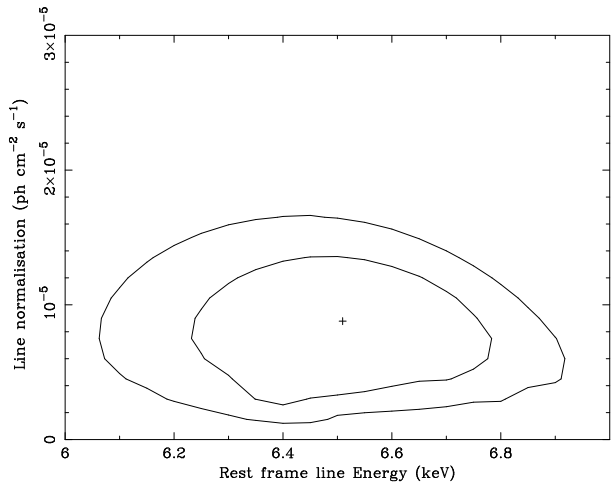}
\caption{Confidence contours (1 and 2 sigma) for the rest frame Fe K
line energy and intensity, following a gaussian fit.}
\label{fig-contfeline}
\end{figure}

\subsection{The 2-12 keV spectrum}

We first fitted the 2-12 keV range with a single power law. We then
checked for the existence of a (redshifted) Fe K line complex around
6-7 keV.  A line was found at rest frame energy of $\sim 6.5$ keV,
with a very poorly defined intrinsic FWHM width of $\sim 0.9 {\rm
keV}$ (see Table~\ref{Tab-params} for the specific values).  The
significance of the line, as measured with the F-test statistic, is
$\sim 98\%$. Confidence contours in the line rest frame energy and
line intensity parameter space are shown in
Fig~\ref{fig-contfeline}. The corresponding equivalent width is $\sim
380^{+230}_{-320}$ eV in the rest frame. These parameters are roughly
consistent with what is expected in Compton-thin type 2 Seyferts,
where multiple components of the Fe K line might be present (see,
e.g., Iwasawa, Fabian \& Matt 1997).

\subsection{The soft excess and photoelectric absorption}

When this fit is  extrapolated  to lower energies, a large soft excess
becomes evident.  We model this by adding a (redshifted) black body,
and a local photoelectric absorption to account both for the Galactic
one and any possible additional absorption in the source (note that we 
expect this to be $\sim 10^{22}\, {\rm cm}^{-2}$ based on the
Balmer decrement of the BLR). The fit results in a $\chi^2=439.93$ for 413
degrees of freedom and a probability of the model not being able to
describe the data of only 83\%.  We tried to fit a plasma emission
model (Raymond-Smith) instead of the black body, but no  good fit
could be achieved.

The rather high black body temperature ($kT_{BB}\sim 140$~eV) suggests
a low black hole mass $M_{BH}\sim 10^4-10^5\, M_{\odot}$, which is
consistent with a luminosity of $\sim 3\times 10^{43}\, {\rm erg}\,
{\rm s}^{-1}$ from such a black hole radiating at near the Eddington
limit.

It is remarkable that the fit to the X-ray data does not require any
photoelectric absorption in excess to the galactic value ($N_{Gal}\sim
1.36\times 10^{20}\, {\rm cm}^{-2}$). In fact we have frozen $N_H$ to
this value and added an extra photoelectric absorption component at
the redshift of the target $z=0.0653$.  Re-fitting the data to this
model finds its best value at no intrinsic absorption with a 3 sigma
upper limit of $1.4 \times 10^{20}\, {\rm cm}^{-2}$. This value is 70
times smaller than the minimum predicted from the BLR Balmer decrement
interpreted in terms of dust reddening and for a normal gas to dust
ratio. It is even 7 times smaller than the value implied by the NLR
reddening, but this might just be due to small covering factor of an
internally reddened NLR.

\begin{figure}
\epsfig{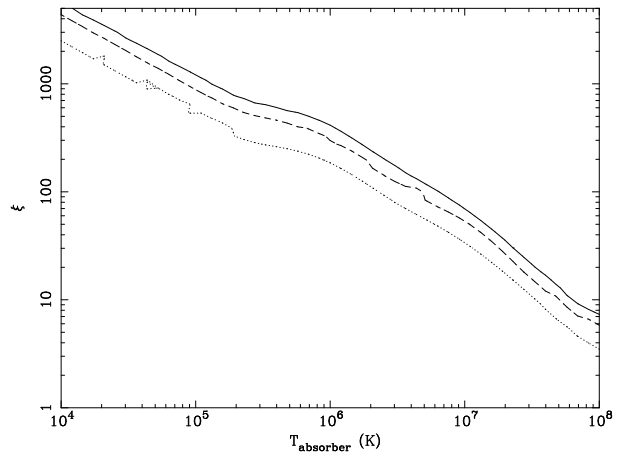}
\caption{Limits for a $N_H=10^{22}\, {\rm cm}^{-2}$ ionized absorber
implied by the EPIC-pn data: 1
sigma (continuous), 2 sigma (dashed) and 3 sigma (dotted).}
\label{noabsori}
\end{figure}

The existence of an ionised absorber is ruled out by the X-ray data.
The addition of a multiplicative {\tt absori} model with a fixed
column density of $N_{H}\sim 10^{22}\, {\rm cm}^{-2}$ does not improve 
the $\chi^2$. Only very large values of the ionisation parameter
$\xi>1000$ or very high temperatures of the absorber ($kT \sim 10^8
\, {\rm K}$) can reach a $\chi^2$ as good as (but not smaller than)
the one without an ionised absorber (see fig~\ref{noabsori}).

\begin{table}
\label{Tab-params}
\begin{tabular}{l c}
Parameter & Value \\
\hline
Photoelectric abs: & {\tt phabs}\\
$N_H$ & $(1.69^{+0.45}_{-0.39})\times 10^{20}\, {\rm cm}^{-2}$\\
Redshifted power law: & {\tt zpowerlw}\\
$\Gamma$ & $1.87\pm 0.05$\\
$A_{\Gamma}$ & $(7.2^{+0.5}_{-0.3})\times 10^{-4}\, {\rm ph}\, {\rm
cm}^{-2}\, {\rm s}^{-1}$\\
Redshifted black body:& {\tt zbbody}\\
$kT$ & $137^{+13}_{-12}\, {\rm eV}$\\
$A_{BB}$ & $(7.1^{+1.4}_{-1.3})\times 10^{-6}$\\
Redshifted Fe K line complex: & {\tt zgaussian}\\
$E_{line}^{\star}$ & $ 6.51\pm 0.30\, {\rm keV}$\\
$\sigma_{line}^{\star}$ & $0.4^{+0.35}_{-0.4}\, {\rm keV}$\\
$F_{line}$ &  $(10.5^{+3.7}_{-4.4})\times 10^{-5}\, {\rm ph}\, {\rm cm}^{-2}\, {\rm
s}^{-1}\, {\rm keV}^{-1}$\\
\hline
\end{tabular}
\caption{Best-fit parameters for the X-ray spectrum of H1320+551
($z=0.0653$ assumed throughout).
Parameters are grouped by model component, and the {\tt xspec} routine
used is also listed. Errors denote the 90\% interval for one parameter
in each case.  The best-fit values and errors have been obtained by
fitting the full 0.2-12.0 X-ray spectrum shown in the text, with the
exception of the parameters labelled with $^{\star}$ which were
obtained by fitting the 2-12 keV spectrum with a redshifted power law
plus a gaussian.}
\end{table}

\subsection{Resulting spectrum}

\begin{figure}
\epsfig{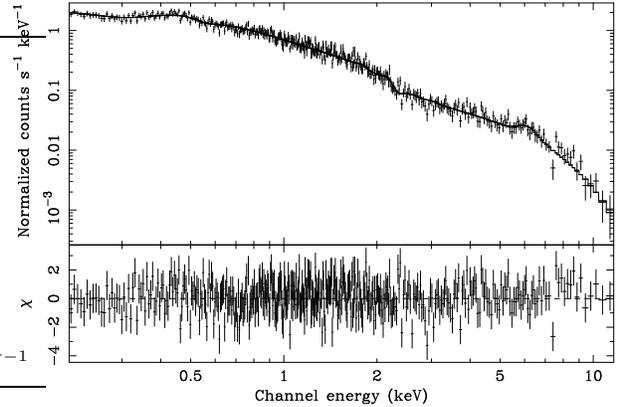}
\caption{EPIC-pn X-ray spectrum together with best fit model (top) and
$\Delta\chi$ of data points to fitted model (bottom).}
\label{Xray-spec}
\end{figure}

Fig \ref{Xray-spec} shows the resulting EPIC-pn spectrum after the
overall fit was performed, together with the contributions to the
$\chi^2$ from each channel. There are no residuals at the
energies of the most prominent absorption edges. The existence of a
slight positive residual at the hard energy end is uncertain as
the background level is very high at these energies.

There is a negative residual at around 0.7 keV in the data (see Fig
\ref{Xray-spec}).  The feature extends for 0.1-0.2 keV with an
amplitude of $\sim 10$ per cent. At the moment, individual calibration
of the EPIC-pn instrument gives residuals of $\sim 3$ per cent or
less, but joint calibration of all XMM-Newton instruments leaves large
discrepancies between EPIC pn and EPIC MOS in the energy range 0.3-1.3
keV of up to 15 per cent. Besides that, a negative residual similar to
ours is seen in many EPIC-pn spectra at around 0.7 keV (S. Sembay,
private communication) and therefore we believe it to be a calibration
artifact.

\begin{figure}
\epsfig{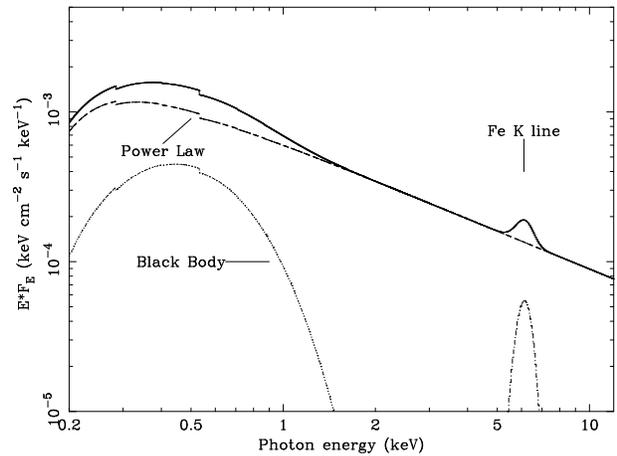}
\caption{Fitted model, in terms of $EF_E$ as a function of energy,
showing the various components.}
\label{Xray-model}
\end{figure}

Fig~\ref{Xray-model} shows the model fitted with the various
components labelled. The flux, corrected for photoelectric absorption
(assumed galactic) of the source is $\sim 1.62\times 10^{-12}\, {\rm
erg}\, {\rm cm}^{-2}\, {\rm s}^{-1}$ in the 0.5-2 keV band and
$2.09\times 10^{-12}\, {\rm erg}\, {\rm cm}^{-2}\, {\rm s}^{-1}$ in
the 2-10 keV band. The luminosity is $3.1\times 10^{43}\, {\rm erg}\,
{\rm s}^{-1}$ in the 0.5-2 keV band and $3.9\times 10^{43}\, {\rm
erg}\, {\rm s}^{-1}$ in the 2-10 keV band.  About 22\% of the 0.5-2
keV luminosity ($\sim 7\times 10^{42}\, {\rm erg}\, {\rm s}^{-1}$) is
contributed by the soft excess that we have modeled as a black body,
which is far too much to be attributable to the host galaxy or to
scatteing.

The fit to the X-ray spectrum is therefore typical
of a Seyfert 1 galaxy. There is an underlying power law, to
which a soft excess, probably the direct quasi-thermal radiation from
the accretion disk, has to be added.  The Fe K emission line complex
is rather broad (FHWM $\sim 900\, {\rm eV}$), as it is often found in Seyfert
2 galaxies, for example in the prototypical NGC 1068 (as a result of
the superposition of various components, Iwasawa et al 1997). The
equivalent width of the complex is $380^{+330}_{-320}\, {\rm eV}$
which lies in between the one expected from a Seyfert 1 and a
Compton-thin Seyfert 2, all consistent with the Seyfert 1.8/1.9 nature of
the source. 

\subsection{Source variability}

\begin{figure}
\epsfig{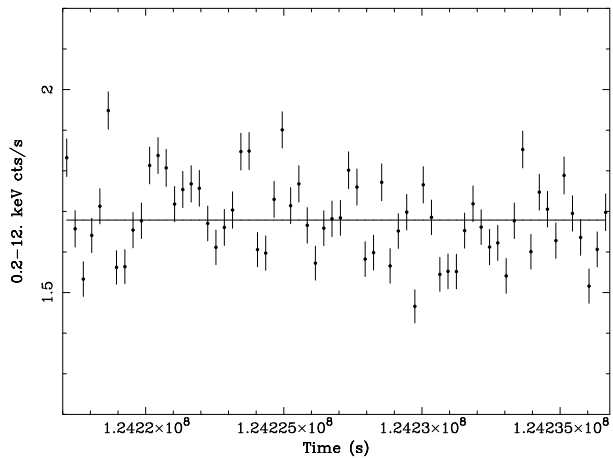}
\caption{EPIC-pn X-ray light curve, binned in 300 s bins. A constant
fit is shown for comparison.}
\label{fig-lightcurve}
\end{figure}

In order to gain further insight into the X-ray properties of this
X-ray source, we have extracted its EPIC pn light curve.  Time
intervals have been binned in 300 s,
resulting in a signal to noise of $\sim 18$. The background counts
have been estimated from the same background subtraction region that
was used to analyze the spectrum. The resulting 0.2-12 keV light curve
is shown in fig~\ref{fig-lightcurve}.

The average count rate is 1.67 ct/s.  However, the light curve is not
consistent with a constant flux. A first glance at
fig~\ref{fig-lightcurve} reveals that 2/3 of the properly computed
1$\sigma$ error bars do not cross the horizontal fit, while only $1/3$
would be expected.  The $\chi^2$ of the fit to a constant intensity is
extremely poor, $\chi^2/\nu = 341/65$.

To estimate the rms variability, we compare the measured variance of
the count rates to the variance expected from the error bars and find
5 per cent intrinsic variability on that time scale.  Although small,
these variations are highly significant. Note that for a $10^5\, {\rm
M}_{\odot}$ black hole, variability is expected down to scales of
seconds.

The fact that the source varies brings additional support to the
rejection of a Compton thick model, which would not predict short term
variability. 

\section{Discussion}

Reconciling the Seyfert 1.8/1.9 character of H1320+551, as drawn from
the optical observation in 1998, with the XMM-Newton data, taken late
2001, has two possible scenarios.  The first one is that the source
spectrum changes over time.  It would then be possible that both the
optical and X-ray absorbing properties change simultaneously. The
other possibility is that both the optical and the X-ray properties
derived from the 1998 optical data and the 2001 XMM-Newton X-ray
observation stay constant over time.  In that case we find that
H1320+551 is not consistent with a Seyfert 1 AGN viewed
through absorbing material.

\subsection{Does the H1320+551 spectrum vary?}

The first issue to address is whether the various data sets give a
consistent picture of H1320+551 or there is rather strong evidence for
varying spectral properties.  As already discussed, the optical
spectra presented here are of superior quality to that of Remillard et
al (1993) and therefore we do not believe there is a strong case for
claiming a change in optical spectral type from Seyfert 1 to Seyfert
1.8/1.9 by comparing the two spectra.

The HEAO-1 data gives a flux that is $\sim 10$ times larger in the
2-10 keV band that any other reported flux in that band.  Although it
is entirely possible that the HEAO-1 flux was correct (therefore
implying a large variation after two decades), this flux might have
been severely contaminated by other sources within the same field of
view of the Modulation Collimator. To further assess this point, we
have searched for RASS sources in the vicinity of H1320+551, and found
a total of 9 sources within a radius of 2 deg, totalling a flux $\sim
5 $ times that of our target.  Although this does not prove that the
HEAO-1 flux is wrong, it illustrates the difficulty in avoiding
confusion problems in the MC-LASS fluxes.

The $ROSAT$ all-sky survey measurement is entirely consistent with the
XMM-Newton one.  The 0.5-2 keV flux measured in the RASS ($(1.9\pm
0.2)\times 10^{-12}\, {\rm erg}\, {\rm cm}^{-2}\, {\rm s}^{-1}$) is
consistent, within errors, with the 0.5-2 keV flux measured in the
EPIC spectrum ($1.6\times 10^{-12}\, {\rm erg}\, {\rm cm}^{-2}\, {\rm
s}^{-1}$). Furthermore, we have folded the best-fit XMM-Netwon model
through the PSPC-B response and found an expected PSPC Hardness Ratio
of $+0.15$, while $0.12\pm 0.10$ was measured. The spectral shape of
H1320+551, as seen in the $ROSAT$ and XMM-Newton observations, are
entirely consistent.

The $ASCA$ data give a 2-10 keV flux very similar to the XMM-Newton
one, but with no soft excess visible. In fact the underlying power law
in the $ASCA$ data is flatter than the XMM-Newton one, with the net
result that, within uncertainties, $ASCA$ finds a harder spectrum than
XMM-Newton.  To further assess that point we simulated the ASCA
spectrum with the model fitted to XMM-Newton.  A single power law with
no absorption gives a fairly acceptable fit ($\chi^2=36.25$ for 26
degrees of freedom) but the power law index is significantly steeper
($\Gamma=2.0\pm 0.1$) than the one measured in the real $ASCA$ data
($\Gamma=1.5\pm 0.1$). That discrepancy looks far too large to be
attributable to cross calibration errors, which are likely to be much
smaller between EPIC-pn and ASCA SIS (Snowden 2002). The fluxes
measured in the fake $ASCA$ data are $1.5$ and $1.7\times 10^{-12}\,
{\rm erg}\, {\rm cm}^{-2}\, {\rm s}^{-1}$ in the 0.5-2 keV and 2-10
keV bands respectively, while $0.7$ and $1.65\times 10^{-12}\, {\rm
erg}\, {\rm cm}^{-2}\, {\rm s}^{-1}$ result from the best power law
fit to the real $ASCA$ data.  While there is agreement in the 2-10 keV
band, the soft excess seen in the XMM-Newton observation, consistently
with the earlier $ROSAT$ observation, is not seen in the $ASCA$ data.

Therefore there is some marginal evidence for X-ray spectral changes
in H1320+551. Whether that implies changing absorption properties is
unclear, and cannot be addressed with the archival $ASCA$ observations
confonted to the XMM-Newton and earlier $ROSAT$ ones.

\subsection{An intrinsic Balmer decrement}

If the WHT and XMM-Newton data presented here are representative of
the true average state of H1320+551, there is an apparently
contradictory behaviour in the optical and in X-rays.  The large
Balmer decrement seen in the BLR is not consistent with the unabsorbed 
X-ray spectrum, if both phenomena have to be explained in terms of
reddening/absorption.

Pappa et al. (2001) have studied a sample of 8 Seyfert 2 galaxies,
where they find at least two without photoelectric absorption in
X-rays.  Three explanations were proposed in that work to account for
the unusual behaviour of these sources: (a) the lack of BLR is real
and intrinsic to the nuclear properties; (b) the Seyfert 2 galaxies
are Compton-thick, in which case the apparent lack of photoelectric
absorption would be due to the $<10$~keV flux coming only from
scattered and/or host galaxy emission and (c) the presence of a dusty
warm absorber reddens the BLR but has little effect in the X-ray
properties. Indeed with the $ASCA$ spectra of these objects, Pappa et
al were unable to find spectral features associated to the warm
absorber.

We have carefully examined these 3 possibilities in the case of
H1320+551.  There are two independent reasons to rule out a
Compton-thick scenario (b).  First we use the three-dimensional
diagnostic diagram proposed by Bassani et al (1999). For H1320+551 the
transmission is $T\sim 1.5$ and the equivalent width of the Fe K line
is $\sim 400$~eV. Both numbers are inconsistent with a Compton-thick
source; instead H1320+551 would be consistently placed in between the
Seyfert 1s, and the Compton-thin Seyfert 2s, which is what would be
expected for a Seyfert 1.8/1.9.  A second independent fact rejecting a 
Compton-thick scenario comes from the detection of X-ray variability
on scales of 300 s.  If the X-rays detected from H1320+551 were due to
scattered radiation, the variability scale would be associated to the
reflector rather than to the nuclear source and it would be much
longer.

The possibility of a dusty warm absorber (c) is also excluded from the
X-ray analysis presented here.  No spectral features are detected in
the X-ray data and in fact any absorbing gas present should be fully
ionized, something rather inconsistent with the presence of dust
reddening the BLR. 

That leaves the intrinsic origin (a) for the large BLR decrement as the
only likely explanation.  This assumption is at odds with the standard
AGN unified model and deserves some further discussion.  In what
follows we address the question on whether H1320+551 is consistent with
an absorbed/reddened type 1 Seyfert or not.

Ward et al (1988) studied the Balmer decrement of the type 1 AGN in
the Piccinotti et al (1982) sample.  The good linear correlation
between Balmer decrement versus the ratio between 2-10 keV luminosity
(mostly unaffected by reddening/absorption) and H$\beta$ luminosity (a
good tracer of absorption) prompted Ward et al (1988) to suggest that
Balmer decrement is determined by nuclear reddening, rather than being
intrinsic to the BLR (see their fig. 4).  They also find an
approximately constant 2-10 keV to H$\alpha$ ratio for the sample (see
their Fig. 5). In fact, Ward et al (1988) conclude that in spite of
the extreme conditions of the BLR, the intrinsic Balmer decrement is
$\sim 3.5$ for the type 1 AGNs.

Now, H1320+551 has a Balmer decrement which is more than half a decade
larger than what would be expected from its 2-10 keV to H$\beta$
ratio.  Reddening correction will not bring this into agreement with
the Seyfert 1s, as both the Balmer decrement and the X-ray to H$\beta$
ratio will decrease if reddening corrected.  On the contrary, the 2-10
keV to H$\alpha$ ratio is entirely consistent with that of the Seyfert
1s.  All that means that the large BLR decrement for this particular
Seyfert 1.8/1.9, together with its unabsorbed X-ray spectrum cannot be 
explained as a Seyfert 1 AGN viewed through obscuring material.

\subsection{Conclusion}

XMM-Newton X-ray observations of the H1320+551, which is classified by
its optical spectrum as a type 1.8/1.9 AGN, reveal no absorption.  If
the non-simultaneous optical and X-ray observations both trace the
true state of this source, we conclude that the large Balmer decrement
of the BLR, which determines its 1.8/1.9 spectroscopic type, is not
due to reddening by dusty absorbing material along the line of
sight. A variety of models can explain a large intrinsic value of the
Balmer decrement, among them the failure of the standard ``case B
recombination'' and/or optically thick BLR clouds. In any case,
the AGN unified model fails completely in this source.

Regardless on whether the unusual optical/X-ray absorption properties
of H1320+551 are due to variations or not, they raise an important
issue for unified AGN models for the X-ray background. H1320+551 is a
source with a relatively soft unabsorbed X-ray spectrum that is
expected to typically have a type 1 AGN optical counterpart.  However,
we identify it with a Seyfert 1.8/1.9, breaking again the one-to-one
identification between X-ray absorption and optical obscuration that
the XRB models use.  Similarly, other relatively soft X-ray sources
with no X-ray absorption might have optical counterparts which deviate
from the standard Seyfert 1 character. If the BLR properties are not
always linked to the absorption displayed by AGN, then Seyfert
1.8/1.9/2 galaxies might appear as optical counterparts of soft X-ray
selected sources as well as type 1 Seyferts often appear as optical
counterparts to hard X-ray sources.

\section*{Acknowledgments}

We are grateful to Steve Sembay and Martin Turner for help with EPIC
calibration issues. The referee is also thanked for
important suggestions on the original version of this paper.  The WHT
telescope is operated on the island of La Palma by the Isaac Newton
Group of Telescopes in the spanish Observatorio del Roque de Los
Muchachos of the Instituto de Astrof\'\i sica de Canarias. Partial
financial support for this work was provided by the Spanish Ministry
of Science and Technology under project AYA2000-1690.

\bsp

\label{lastpage}

\end{document}